\documentstyle[aps,twocolumn,floats,prl]{revtex}
\input psfig

\begin{document}
\draft
\title{Comment on the Appropriate Null Hypothesis for Cosmological
Birefringence}
\author{Daniel J.\ Eisenstein}
\address{Institute for Advanced Study, Olden Lane, Princeton, NJ 08540;
eisenste@sns.ias.edu}
\author{Emory F.\ Bunn}
\address{Physics and Astronomy Department, Bates College, Lewiston, ME 04240;
ebunn@abacus.bates.edu}
\maketitle
\thispagestyle{empty}

A recent {\it Letter} \cite{nod97} claims to have discovered evidence
for birefringence in the propagation of radio waves across cosmological
distances.  Unfortunately, this claim is based on a flawed statistical
analysis.

To search for birefringence, the authors look for correlations between
the direction and distance to a galaxy and the angle $\beta$ between the 
polarization direction and the galaxy's major axis.
Plotting the data as shown in Fig.\ 1(d) of their paper, here
reproduced as Fig.\ 1, they use the correlation coefficient $R_{xy}$
as their statistic.

To estimate the significance of their result, the authors use mock data
samples constructed by randomly picking the angle $\beta$ from a uniform
distribution of allowed angles\footnote{To be specific, they choose
from random distributions the directions of each galaxy's major axis
and plane of polarization.  This results in a uniform distribution (in
the 1st and 3rd quadrants) for
$\beta$, the angle between these two directions.}.  This is not the
proper null hypothesis for testing the dependence of $\beta$ on the
direction and distance to the galaxies.  Rather, for the null
hypothesis, one should draw the angles from the observed distribution,
which in the case of the high-redshift subsample, from which the
primary conclusions were drawn, clearly is not a uniform distribution.
From Fig.\ 1, one can see by eye that the polarization in these
galaxies tends to align with the galaxy minor axis, i.e.\ $\beta^\pm$
prefers $\pm\pi/2$ and avoids $0$ or $\pm\pi$.  For example, $\sim3/4$
of the points have $\pi/4\le|\beta| \le 3\pi/4$.

That this matters can be easily seen from the following example.
Consider the region in the $x$-$y$ coordinate plane spanning $-1$ to 1 in
both directions.  If we uniformly fill the first and third quadrants,
the correlation coefficient $R_{xy}$ will be 0.75.  If, however, we fix
$y= 0.5\,\rm{sgn}(x)$ while allowing $x$ to span $-1$ to 1 uniformly as
before, then $R_{xy}= \sqrt{3}/2 \approx 0.867$.  Collapsing the $y$
direction in this way allows more of the scatter to be explained by 
the best-fit line.

Hence, we should expect that the tendency of the angle $\beta$ to
prefer $\pm \pi/2$ will cause $R_{xy}$ to be higher than it would be if
$\beta$ were uniformly distributed between 0 and $\pm \pi$.  By using
the latter as their null hypothesis, the authors find a spuriously high
statistical significance for their result.  Indeed, if the underlying
galaxy population truly had a uniform intrinsic distribution of
$\beta$, it would be impossible to measure the proposed birefringence
at all; one could not detect a rotation of such a distribution.

\begin{figure}
\centerline{\psfig{file=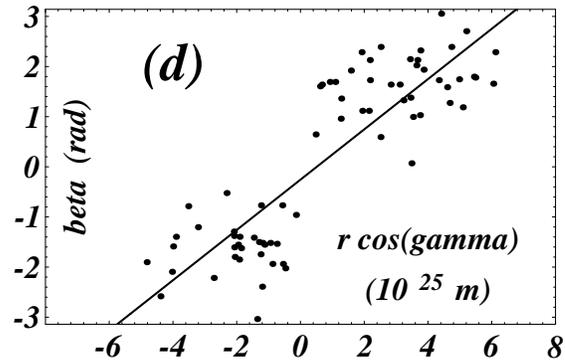}}
\caption{Figure 1(d) reproduced from [1].
71 galaxies with redshifts above 0.3 are shown.  $\gamma$
is the angle from the proposed birefringence direction, $r$ is the
distance to the galaxy, and $\beta$ is the angle between the galactic
major axis and the polarization direction.  See [1] for more details.
Our claim is that the $\beta$ are not uniformly distributed between 0
and $\pm\pi$ but rather are clumped toward $\pm\pi/2$.
}
\label{figa}
\end{figure}

Stated another way and estimating by eye, in Fig.\ \ref{figa} the data
{\it are} more tightly correlated than they would be if the $\beta$
values were randomly and uniformly distributed between $0$ and $\pm
\pi$.  However, they are not significantly more correlated than they
would be if the $\beta$ values in a quadrant were shuffled among
themselves while the best-fit line was adjusted accordingly.  Hence,
the claimed correlation of the angle $\beta$ with the position and
distance of the galaxy is not statistically significant.

Taking the null hypothesis that the birefringence does not exist and that
the angles between the polarization directions and galaxies' major axes
are distributed as the data indicate, one is left to explain why the
particular direction in the sky turned out to yield a higher $R_{xy}$
than other directions.  This most likely results from combining the
inhomogeneous sky coverage---the sample is mostly from the Northern sky
and avoids low galactic latitudes---and the propensity of the chosen
statistic $R_{xy}$ to prefer directions that place many galaxies near
the center of the spread in $r \cos{\gamma}$ where the tendency of
$\beta$ to prefer the center of its range can best reduce the scatter.
It is not surprising that such a direction could exist.

A second error relating to the choice of statistic and null hypothesis
is the authors' use of the slope of the best fit line in Fig.\ \ref{figa} 
as a measure of the inverse birefringence scale $\Lambda_s^{-1}$.
Because the null hypothesis (either the one they used or the one proposed
here) produces a non-zero slope in the absence of birefringence, this
is clearly a highly biased and inappropriate estimator.

D.J.E.\ was supported by NSF PHY-9513835.

\clearpage

\pagestyle{empty}
\centerline{{\large\bf Reply to astro-ph/9705190}}
\bigskip
\bigskip

Nodland \& Ralston [1] claim to have discovered evidence
for birefringence in the propagation of radio waves through the
universe.  We have argued [2] that flaws in their statistical analysis
decrease the statistical significance of their results; Carroll
\& Field [3] reach similar conclusions.  Further data from Leahy [4]
and Wardle et al.\ [5] strongly argue against the proposed effect.
In [6], Nodland \& Ralston responded to our critique.  Here, we 
explain why their response is incorrect.

Nodland \& Ralston base their reply [6] to our Comment on claims that
several of our assertions were in error.  They raise a number of
different issues, but the chief arguments appear to fall into
three categories.  They claim 1) that the distribution of the angle
$\beta$ used in their Monte Carlo methods differs from the uniform
distribution we ascribed to it, and 2) that one could measure a
birefringence even if the underlying angular distribution(s) were
uniform.  They also claim 3) that our methods would obscure a
reasonable linear correlation even if it were in the data.  We show
here in detail that they are in error in these three claims.  Indeed,
we find nothing in their reply that gives us reason to modify our
Comment.

\bigskip
1) Nodland \& Ralston object to our claim that the angle $\beta$ in
their simulations is drawn from a uniform distribution.  We show here
that our assertion is correct, i.e.\ that the simulated $\beta$'s
are indeed uniformly distributed.

The angle used in their statistical analysis does indeed depend on the
galaxy's position angle $\psi$, its polarization angle $\chi$,
its position on the sky, and the proposed birefringence direction
$\vec{s}$.  However, the analysis procedure splits into two pieces: 1) the
determination of the angles $\beta^+$ and $\beta^-$ from $\chi$ and
$\psi$, and 2) the choice of which of these angles to use given the 
particular relative position of the galaxy and the birefringence
direction (i.e. $\beta^+$ in one half-space and $\beta^-$ in the
other).  The search over many different $\vec{s}$, as detailed in their
procedures, never alters the values of $\beta^+$ or $\beta^-$; it
merely alters which one of them is used at any given time.  Hence,
once $\beta^+$ and $\beta^-$ are determined from $\chi$ and $\psi$
at the beginning of the calculation, one need never refer to $\chi$
and $\psi$ again.

Moreover, the definitions of $\beta^+$ and $\beta^-$ reveal that
$\beta^-$ = $\beta^+ - \pi$.  Hence, knowledge of $\beta^+$ for
each galaxy at the beginning of the calculation, before calculating
the correlation coefficient with respect to any birefringence direction, 
is all the information needed about the galaxy's 
polarization angles and position angles.  This is true for both
of the Monte Carlo procedures described in [1].

Hence, to prove that our treatment is equivalent to theirs, all we need
do is show that drawing $\chi$ and $\psi$ from independent uniform
distributions yields a uniform distribution of $\beta^+$.  This is
fairly clear, but for completeness, we now show the algebra.

Assume that for a given galaxy the angles $\chi$ and $\psi$ are drawn
from independent uniform distributions on $[0,\pi]$.  Now, consider the
quantity $\Delta\equiv\chi-\psi$.  The probability density for this
quantity is obtained by convolving the uniform distributions of $\chi$
and $\psi$.  The result is a triangle-shaped distribution,
\begin{equation}
f_\Delta(\Delta)={1\over\pi^2}(\pi-|\Delta|),
\end{equation}
for all values of $\Delta$ between $-\pi$ and $\pi$.
(Here $f_\Delta$ is an ordinary probability density: the probability
that $\Delta$ lies between $\hat\Delta$ and $\hat\Delta+ d\hat\Delta$
is $f_\Delta(\hat\Delta)\,d\hat\Delta$.)
Now apply equation (2) of [1] to convert $\Delta$ to
$\beta^+$.  Given any $\beta^+$ between 0 and $\pi$, there
are two values of $\Delta$ that can correspond to this $\beta^+$:
either $\Delta=\beta^+$ or $\Delta=\beta^+-\pi$.  Therefore, the
probability density for $\beta^+$ is
\begin{eqnarray}
f_{\beta^+}(\beta^+) &=&f_\Delta(\beta^+)+f_\Delta(\beta^+-\pi)\\
&=& {1\over\pi^2}(\pi-\beta^+)+{1\over\pi^2}(\pi-(\pi-\beta^+))={1\over\pi}.
\end{eqnarray}
So $\beta^+$ is uniformly distributed on $[0,\pi]$.
Hence, Nodland \& Ralston's claim that our treatment differs from theirs is
incorrect.

Real radio galaxies do not have uniformly distributed values of
$\beta$.  This is clear from Figure 1d of [1], in which
$|\beta^{\pm}|$ is clearly seen to cluster around $\pm\pi/2$, and it is also
well known from previous studies of radio galaxies.  Our argument is
that the incorrect assumption of uniformity artificially inflates the
statistical significance of their results.

Nodland \& Ralston defend their procedure by pointing out that $\chi$ and
$\psi$ are uniformly distributed in the real data.  That is correct but
irrelevant.  Only differences between $\chi$ and $\psi$ enter the
calculation, and the {\it correlation} between the two angles causes
these differences (encoded in $\beta^\pm$) to be distributed
nonuniformly.  By drawing $\chi$ and $\psi$ from {\it independent}
uniform distributions, Nodland \& Ralston fail to take this correlation
into account, leading to incorrect results.

\bigskip
2) Nodland \& Ralston dispute our statement that ``if the underlying
galaxy population truly had a uniform distribution of $\beta$, it
would be impossible to measure the proposed birefringence at all.''
To see why our statement is true, remember that one is trying to
measure an {\it additional} path-length-dependent variation on top of
the intrinsic angle between the polarization and the major axis.  If
the polarization direction as the radiation left the galaxy were
unrelated to the galaxy position angle $\psi$, one would never be able
to distinguish the initial polarization angle from the rotation induced
as the light traveled to us.  Only by knowing something about the
relation between the intrinsic polarization direction and some other
observable property of the galaxy can one measure an additional
path-length-dependent rotation.

For a mathematical treatment of this, let us suppose for the moment
that the direction $\vec s$ of the axis is fixed.  Then as we have
argued, the Nodland-Ralston null hypothesis is equivalent to drawing
each $\beta^+$ value from a uniform distribution, generating $\beta^-$,
and then choosing the appropriate one according to the sign of
$\cos\gamma$.  Now, suppose that we ``rotate'' each $\beta$ in this
data set by adding (modulo $\pi$) the amount ${1\over
2}\Lambda_s^{-1}r\cos\gamma$ to each $\beta$.  Since the original
$\beta$'s were independently and uniformly distributed, the resulting
$\beta$'s will have precisely the same statistical distribution as
before: they will be independent and uniformly distributed.  There is
therefore no statistical test that can distinguish between the
``rotated'' and ``unrotated'' data sets.

Nodland \& Ralston propose as a counterexample a data set in which
there is a perfect linear relation between $\beta$ and $r\cos\gamma$.
While this data set has a uniform distribution of the observed
$\beta$, it is not a case in which the {\it underlying} galaxy
population satisfies this uniform distribution, and hence it has no
bearing on the question at hand.  Monte Carlo data sets drawn from
uniform intrinsic distributions of $\beta$, with or without an
additional ${1\over 2}\Lambda_s^{-1}r\cos\gamma$ rotation, would have
a negligible probability to place all of the points so nicely on a
straight line.  Since none of the Monte Carlo sets would look anything
like the data (as quantified, say, by the correlation coefficient),
this data would be not only striking evidence for birefringence {\it
but also} evidence against an uniform distribution of intrinsic
$\beta$'s.  Indeed, it is exactly because the birefringent model
would reduce the $\beta$ distribution to a distance-independent
distribution that one favors this interpretation.  Hence the
counterexample does not bear upon our claim, as it does not satisfy
the supposition of our assertion.

\bigskip
3) Nodland \& Ralston claim that shuffling the data would fail to
detect a perfectly correlated $\delta(y-x)$ distribution.  This is 
incorrect.  Shuffling means randomly matching the $x$ coordinate of 
one data point with the $y$ coordinate of another.  If the data lay
on an inclined straight line, no permutations of the data would ever
produce a data set with as high a correlation; the new data sets would
generically show large amounts of scatter.  In this particular
case, {\it none} of the shuffled data sets would have as high a
correlation as the original data, since the latter has the maximum
possible correlation coefficient ($r=1$).
Hence, the correlation would be detected at high significance.

\bigskip
In the latter half of their reply, Nodland \& Ralston merely restate
our case.  When they performed a somewhat more correct statistical
procedure (shuffling the $\beta$'s), the statistical significance of
their result dropped considerably.  They claim the signal is still
significant, but they do not account for the fact that one has looked at
many possible directions in the sky.  In short, they confirm our
essential point that using the distribution of $\beta$'s from the
observed distribution increases the correlation measured in the
non-refringent universe, thereby decreasing the statistical significance of
their claims.  A similar calculation may be found in Carroll \& Field
[3, p.\ 10].

In conclusion, we find that none of the arguments in [6] provide
us with any reason to modify the conclusions of our original Comment.

\end{document}